\documentclass{procstyle}

\usepackage{refmerge} 

\newcommand{\fzero}{\ensuremath{f_0}(980)\xspace}

\newcommand{\bei}{\begin{itemize}}
\newcommand{\eei}{\end{itemize}}
\newcommand{\beq}{\begin{equation}}
\newcommand{\eeq}{\end{equation}}
\newcommand{\beqn}{\begin{eqnarray}}
\newcommand{\eeqn}{\end{eqnarray}}
\newcommand{\beqns}{\begin{eqnarray*}}
\newcommand{\eeqns}{\end{eqnarray*}}

\newcommand{\mc}{\multicolumn}

\RequirePackage{xspace}
\usepackage{relsize}
\def\babar{\mbox{\slshape B\kern-0.1em{\smaller A}\kern-0.1em
    B\kern-0.1em{\smaller A\kern-0.2em R}}}

\def\ea{{\it et al.}}
\def\nut{$\,\nu_\tau$}
\def\piz{$ \pi^0 $}

\def\eg{{\it e.g.}}

\newcommand\ph{\phantom}
\def\rs{\raisebox{1.5ex}[-1.5ex]}

\input myBabarsym

\begin{document}

\title{\boldmath 
	The Measurement of Time-Dependent \CP-Violating Asymmetries 
	in Loop-Dominated \B Decays with \babar}

\author{Andreas H\"ocker \\ 
	{\lowercase{(representing the \babar\  \uppercase{C}ollaboration)}}}

\address{Laboratoire de l'Acc\'el\'erateur Lin\'eaire,\\
         IN2P3-CNRS et Universit\'e Paris-Sud --
         B\^at. 200, BP34 -- F-91898 Orsay, France\\
         E-mail: hoecker@lal.in2p3.fr}

\twocolumn[\maketitle\abstract{
We report on preliminary measurements of time-dependent \CP asymmetries 
in neutral \B decays to \CP eigenstates with transition amplitudes that 
are dominated by penguin-type loops. The results are obtained from a 
data sample of up to $227$ million $\FourS \to B\Bbar$ decays collected 
with the \babar\  detector at the \pep2 asymmetric-energy $B$-meson 
Factory at SLAC. The amplitudes of the effective mixing-induced \CP asymmetries, 
$\stwob_{\rm eff}$, are derived from decay-time distributions of
events in which one neutral \B meson is fully reconstructed in a final 
state without charm and the other \B meson is determined to be either 
a \Bz or \Bzb from its decay products.
}]

%
%

\section{Introduction}

Since the precise measurement of \stwob from the
time-dependent \CP-asymmetry analyses of $b\to c\cbar s$ 
decays\cite{sin2bbabar,sin2bbelle}, considerable effort is spent 
at the \B-Factory experiments to determine the corresponding 
quantity from penguin-loop-dominated (charmless) \B decays 
(``$s$-penguin decays''), such as $\Bz\to\phi\Kz$, $\etapr K^0$, 
$\fzero K^0$, $S$-wave dominated $K^+K^-K^0$ and $\piz\Kz$.
Due to the large virtual mass scales occurring in the penguin loops,
additional diagrams with heavy particles in the loops and new
\CP-violating phases may contribute to the decay amplitudes.
The measurement of \CP violation in these channels and the 
comparison with the charmonium reference value is therefore a 
sensitive probe for physics beyond the Standard Model (SM).
Due to the different non-perturbative strong-interaction properties 
of the various $s$-penguin decays, the NP-induced deviation from 
\stwob is expected to be channel-dependent.

There is a variety of reasons why specific contributions of New 
Physics (NP) are expected at or below the TeV scale.
Among these are the gauge hierarchy problem of the 
SM Higgs sector, Baryogenesis, Grand Unification of the gauge 
couplings and the strong \CP problem. If the NP is not flavor 
blind it should be present in the flavor sector, and detectable 
in loop-induced \B decays.\cite{NewPhys1,NewPhys2}
However, the present insignificance of NP effects already sets 
strong boundary conditions on the generic flavor structure of 
the NP, hence favoring minimal flavor-violating schemes.\cite{grossman}

Assuming penguin dominance and neglecting CKM-suppressed amplitudes,
$s$-penguin decays carry approximately the same weak phase as the 
decay $\Bz\to\jpsi\KS$. Their mixing-induced \CP-violating parameters 
are expected to be\cite{NewPhys1,phases} $-\eta_f\times\stwob\simeq-\eta_f\times0.7$ 
in the SM\footnote
{
	With these assumptions, the expected phase shift is
	$2\beta_{\rm eff}-2\beta=
	-\arg[(V_{tb}V_{ts}^*V_{cb}^*V_{cs})^2]\simeq2.1^\circ$.
}, 
where $\eta_f$ is the \CP eigenvalue of the final state $f$. 

This simple relation is altered by the presence of a suppressed 
$V_{us}V_{ub}^*$ penguin diagram in all $s$-penguin decays, and a
corresponding color-suppressed tree-level diagram in three-body 
decays (such as $\Kp\Km\Kz$) or resonant decays with non-$s\sbar$
flavor contributions (like $\etapr\Kz$, $f_0\Kz$ and $\piz\Kz$). As a 
consequence, only an effective $\stwob_{\rm eff}=-\eta_f\times S$ 
is determined.
Naive dimensional arguments may allow us to derive coarse estimates 
for the mode-specific theoretical systematics on the amplitude level. 
We find an uncertainty of ${\cal O}(5\%)$ (corresponding to a $2\beta_{\rm eff}$ shift 
of about $2^\circ$) for the golden mode $\phi\Kz$, ${\cal O}(10\%)$ for the 
silver-plated modes $\etapr\Kz$, $f_0\Kz$ or $\Kp\Km\Kz$, and  
${\cal O}(20\%)$ for the bronze-plated modes that have no $s\sbar$ 
component ($\piz\Kz$ and similar). Computations based on QCD 
factorization generally find smaller deviations.\cite{BN} 
Bounds on $2\beta_{\rm eff}-2\beta$ can be derived from 
SU(3) flavor-symmetry relations,\cite{ligetiquinn} which are however 
weak at present.

We present in this summary preliminary measurements of \CP-violating asymmetries 
in all of the above mentioned $s$-penguin decays. The data used for these
measurements were accumulated with the \babar\  detector\cite{bib:babarNim} 
at the \pep2 asymmetric-energy $e^+e^-$ storage ring at SLAC. The sample
consists of $227\times10^{6}$ $\B\Bbar$ pairs ($209\times10^{6}$ for 
$\fzero\KS$) collected at the 
\FourS resonance. Detailed descriptions of the individual analyses 
presented here are given in Refs.\cite{babarall}.
In Ref.\cite{bib:babarNim} we describe the silicon vertex tracker 
(SVT) and drift chamber used for track and vertex reconstruction, the 
Cherenkov detector (DIRC), the electromagnetic calorimeter (EMC), 
and the instrumented flux return (IFR).

With $\deltat \equiv t_{\CP} - t_{\rm tag}$ defined as the proper 
time interval between the decay of the fully reconstructed $B^0_{\CP}$ 
and that of the  other meson $\Bz_{\rm tag}$,  the time-dependent decay
rate is given by 
\beqn
\label{eq:theTime}
f_{{\rm Q_{tag}}}(\deltat) & = & 
        \frac{e^{-\left|\deltat\right|/\tau}}{4\tau}
        \bigg( 1 + {\rm Q_{tag}} S\sin(\deltamd\deltat) \nonumber \\
  &&	 -\;{\rm Q_{tag}} C\cos(\deltamd\deltat)
        \bigg)\,,
\eeqn
where   $Q_{\rm tag}= 1(-1)$ when the tagging meson $\Bz_{\rm tag}$
is a $\Bz(\Bzb)$, $\tau$ is the mean \Bz lifetime, and $\deltamd$ is 
the $\BzBzb$ oscillation frequency. The parameter $S$ is non-zero if 
there is mixing-induced \CP violation, while a non-zero value for $C$
would indicate direct \CP violation. At least two interfering decay 
amplitudes with different weak and strong phases are required to obtain 
$C\ne0$. Hence we do not expect to observe significant direct \CP 
violation in any of the channels studied if the assumption of penguin 
dominance is correct.

%
%

\section{Analysis Techniques}
\label{sec:dataAnalysis}

We reconstruct signal candidates from combinations 
of a $\KS$ candidate decaying to $\pi^+\pi^-$, and a
$\phi\to\KpKm$, $\etapr\to(\eta\pip\pim,\,\rho^0\gamma)$,
$\fzero\to\pip\pim$, $\Kp\Km$ (excluding $\phi$), 
or a $\piz$ candidate.
We also reconstruct the mode $\Bz\to\phi\KL$, where a
$\KL$ candidate is identified either as an unassociated 
cluster of energy in the EMC or as a cluster of hits in 
the IFR.
For the $\Bz\to\etapr\KS$ decay, also $\KS\to\piz\piz$
and $\eta\to(\gamma\gamma,\,\pip\pim\piz)$ decays 
are considered so that the reconstructed signal amounts
to $46\%$ (not containing the detector acceptance) of 
the total $\etapr\KS$ production.

We use strong particle identification from the DIRC, the 
tracking system and the EMC, to identify pions and kaons 
in the final state. Two kinematic variables are used to 
discriminate between signal-$B$ decays and combinatorial 
background: the difference $\DeltaE$ between the center-of-mass 
(CM) energy of the \B~candidate and the CM beam energy, and the 
beam-energy-substituted \B-candidate mass $\mes$, defined 
in the laboratory frame.  
In the $\piz\KS$ analysis, these variables are replaced by
the unconstrained \B-candidate mass $m_{\rm rec}=|q_B|$, and the 
missing mass $m_{\rm miss}=|q_{\epem}-\hat q_{B}|$, where 
$q_{\epem}$ is the four-momentum of the $\epem$ system 
and $\hat q_B$ is the four-momentum of the \B candidate 
after applying a \Bz-mass constraint. The missing mass has
a slightly better resolution than $\mes$ and, by 
construction, $m_{\rm rec}$ and $m_{\rm miss}$ have a 
vanishing linear correlation coefficient.
%

Continuum $e^+e^-\to q\bar{q}$ ($q = u,d,s,c$) events are the 
dominant background. To enhance discrimination between signal 
and continuum, we use a multivariate analyzer (MVA), which is a
Fisher discriminant or a neural network (NN), to combine four 
discriminating variables: the cosine of the angle between the 
CM $B$ direction and the $z$ axis (along the beam direction), 
the cosine of the angle 
between the thrust axis of the $B$ candidate and the $z$ axis, 
and the topological monomials $L_0$, $L_2$ (see Refs.\cite{babarall}). 

Monte Carlo (MC)-simulated events are used to study the background 
from other \B decays. The charmless modes are grouped into 
different classes with similar kinematic and topological 
properties. In the $\Kp\Km\KS$ analysis exclusive $b\to c$ 
decays with the same final state as the signal are vetoed 
by cuts on their invariant masses.

The time difference $\deltat$ is obtained from the measured 
distance between the $z$ positions
of the $\Bz_{\CP}$ and $\Bz_{\rm tag}$ decay vertices, and the 
boost $\beta\gamma=0.56$ of the \epem\  system. Due to the 
$\KS$ lifetime, the $\deltat$ determination is particularly 
challenging for the $\piz\KS$ final state. To select valid 
candidates for this mode, at least 4 SVT hits are required
for each of the decay pions (about $60\%$ of the events). 
While the $\deltat$ information of the rejected events is not 
used, they contribute to the measurement of the $C$ coefficient. 
We determine $\deltat$ from 
a global constrained fit to the entire $\FourS\to\Bz\Bzb$ 
decay tree that takes the information on the beam energy
and the position of the interaction point into account. To
further improve the $\deltat$ resolution, the sum of the two 
measured \Bz lifetimes is constrained to $2\tau$ with an 
uncertainty of $\sqrt{2}\tau$ in the fit.

A new NN-based algorithm\cite{newtag} has 
been developed to determine the flavor of the 
$\Bz_{\rm tag}$. In addition to the information exploited 
by the algorithm described in Ref.\cite{sin2bbabar}, it uses 
low-momentum electrons, $\Lambda\to p\pi$ decays, and 
additional correlations among identified kaon candidates.
The resulting effective tagging efficiency amounts to $Q\simeq30.5\%$.
The tagging efficiencies, mistag probabilities,
mistag biases and also the $\deltat$ resolution parameters
are determined from a fit to fully reconstructed \B decays to 
charm.

We use unbinned extended maximum-likelihood fits to extract 
the event yields and the \CP parameters defined in Eq.~(\ref{eq:theTime}). 
In these fits, the likelihood for a given event is 
the sum of the signal, continuum and the $B$-background 
likelihoods, weighed by their respective event yields. We use 
as fit variables $\deltat$, $\DeltaE$, $\mes$, the 
MVA (all modes) as well as the resonance mass and decay angle 
($\phi\KS$ and $\fzero\KS$). In the case of $\phi\KL$ the \B-mass kinematic 
constraint is necessary to determine the $\KL$ momentum so that
$\mes$ cannot be exploited in the fit. The signal reference shapes 
for the fit variables are obtained from MC simulation,
validated with control samples from fully 
reconstructed \B decays to charm of similar topology.
As many background shape parameters as possible are simultaneously 
determined by the fit to reduce the use of prior assumptions.

The $\Kp\Km\KS$ final state has unknown a priori \CP content so 
that the \CP-even fraction, $f_{\rm even}$, has to be extracted 
experimentally. Two different methods are applied, of which the 
first one (described below) provides the main result, and the second 
is a cross check. We study the $\Kp\Km$ helicity angle distribution 
$|A_{\Bz\to\Kp\Km\KS}(\cos\theta)|^2=\sum_{\ell=0,1,2}\langle P_\ell\rangle
\times P_\ell(\cos\theta)$, where $P_\ell$ are Legendre polynomials
of order $\ell$. The average moments $\langle P_\ell\rangle$ are
computed by means of background-subtracted signal weights returned
by the fit.\cite{splots} We find
$f_{\rm even}=A_S^2/(A_S^2+A_P^2)=0.89\pm0.08\pm0.06$, where the
first error is statistical and the second systematic, and
where the $S$- and $P$-wave amplitude-squared are given by 
$A_S^2=\sqrt{2}\langle P_0\rangle-\sqrt{5/2}\langle P_2\rangle$
and 
$A_P^2=\sqrt{5/2}\langle P_2\rangle$. Assuming isospin invariance,
$f_{\rm even}$ can also be obtained from the ratio\cite{belleiso}
$2\Gamma(\Bp\to\Kp\KS\KS)/\Gamma(\Bz\to\Kp\Km\KS)$, for which we
find $0.75\pm0.11$ (statistical error only), in agreement with the
result from the moments analysis.

\section{Results}

\begin{table*}[t]
\begin{center}
\setlength{\tabcolsep}{0.0pc}
\begin{tabular*}{\textwidth}{@{\extracolsep{\fill}}lrcc}\hline
&&&\\[-0.3cm]
Mode    	& \mc{1}{r}{Yield~~~}
				& $\stwob_{\rm eff}$	&	$C$ \\[0.1cm]
\hline
&&&\\[-0.3cm]
		& {\small$(\phi\KS)$} $114\pm12$ 	
				& 			& \\
\rs{$B\to\phi\Kz$}	
		& {\small$(\phi\KL)$} $\ph{1}98\pm18$ 	
				& \rs{$+0.50\pm0.25^{\,+0.07}_{\,-0.04}$}
							& \rs{$\ph{+}0.00\pm0.23\pm0.05$}\\
$B\to\etapr\KS$	& $819\pm38$	& $+0.27\pm0.14\pm0.03$
							& $-0.21\pm0.10\pm0.03$\\
$B\to\fzero\KS$	& $152\pm19$	& $+0.95^{\,+0.23}_{\,-0.32}\pm0.10$
							& $-0.24\pm0.31\pm0.15$\\
$B\to\Kp\Km\KS$	& $452\pm28$	& $+0.55\pm0.22\pm0.04\pm0.11_{\CP}$
							& $+0.10\pm0.14\pm0.06$\\
$B\to\piz\KS$	& $300\pm23$	& $+0.35^{\,+0.30}_{\,-0.33}\pm0.04$
							& $+0.06\pm0.18\pm0.06$\\[0.1cm]
\hline
\end{tabular*}
\caption[.]{\label{tab:results}
	Preliminary fit results for the different $s$-penguin modes. 
	The second column gives the observed signal yield, the third 
	the effective $\stwob_{\rm eff}=-\eta_f\times S$ value, and the last 
	column the result for the $C$ parameter. The first errors given are
	statistical and the second systematic. The third error given for 
	$\stwob_{\rm eff}(\Kp\Km\KS)$ is due to the uncertainty on the 
	\CP-even fraction.
	}
\end{center}
\end{table*}$\!\!$
The fit results for the event yields and \CP parameters are given 
in Table~\ref{tab:results}. In the case of $\Kp\Km\KS$, 
$\stwob_{\rm eff}$ and $f_{\rm even}$ (approximately related through
$\stwob_{\rm eff}\sim S/(2f_{\rm even}-1)$) are simultaneously 
determined by the fit. The individual $\stwob_{\rm eff}$ results 
for the $\phi\KS$ ($0.29\pm0.31$) and $\phi\KL$ ($1.05\pm0.51$) 
subsamples agree within statistical errors.
We have also measured the charge asymmetry
in the decay $\Bp\to\phi\Kp$ for which we find the null result
$A_{\phi\Kp}=(\Gamma(\phi\Km)-\Gamma(\phi\Kp))/\Sigma=+0.054\pm0.056\pm0.012$.

All measurements reported here are statistics limited. The systematic 
errors are dominated by the opposite \CP background and the reference 
shape modeling ($\phi\Kz$), the fit bias ($\etapr\KS$, $\fzero\KS$ and 
$\Kp\Km\KS$), the interference with other resonances ($\fzero\KS$), 
and the background tagging asymmetries and SVT alignment ($\piz\KS$).
All measurements have in common the uncertainty due to the
unknown \CP violation on the tag-side.

The individual $\stwob_{\rm eff}$ results are in agreement with the 
charmonium value,\cite{newtag} with the exception of $\etapr\KS$,
which exhibits a $3.0\sigma$ discrepancy (neglecting theoretically
motivated deviations). Overall, the results are in agreement with but 
more precise than the corresponding numbers from the Belle 
Collaboration\cite{bellespeng}.

\begin{figure}[t]
\centerline{\psfig{file=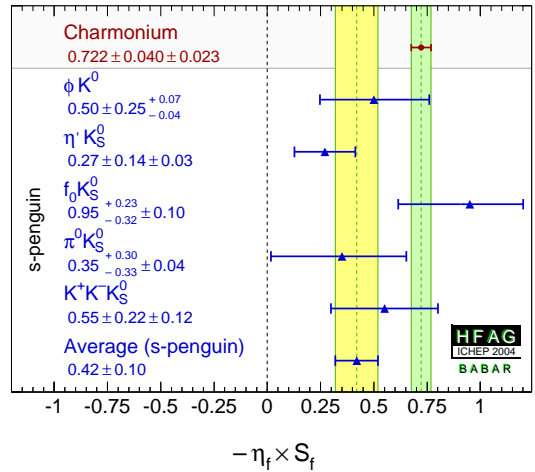,width=70mm}}
\vspace{-0.3cm}
\caption[.]{\label{fig:hfag}
	Compilation of preliminary results on $\stwob_{\rm (eff)}$ from 
	charmonium and $s$-penguin decays. The averages, indicated by 
	the shaded bands, are obtained ignoring theoretical uncertainties.
	}
\end{figure}

A graphical compilation of the results is 
given in Fig.~\ref{fig:hfag}. We find the $s$-penguin average 
$\stwob_{\rm eff}=0.42\pm0.10$ (${\rm C.L.}=0.40$), where theoretical 
uncertainties are ignored. The difference with the charmonium 
reference amounts to $2.7$ standard deviations ($\sigma$). Adding the 
aforementioned individual theoretical uncertainties, and treating
them as allowed ranges,\cite{rfit} the significance of the 
discrepancy drops to $2.0\sigma$. 

None of the modes studied exhibits non-zero direct \CP violation. 
The $s$-penguin average is $C=-0.081\pm0.073$ (${\rm C.L.}=0.43$).

\section{Summary}

We have presented preliminary results on mixing-induced and direct \CP 
violation for penguin-loop-dominated \B decays, using the full 
Summer 2004 data sample. We have observed\cite{fzeroks} and updated 
the new silver-plated mode $\Bz\to\fzero\KS$. Sophisticated
vertexing\cite{pizeroks} allows us to measure the mixing-induced
\CP asymmetry in $\Bz\to\piz\KS$, using the $\KS$ decay into 
$\pip\pim$. With the exception of $\Bz\to\etapr\KS$, 
all individual $s$-penguin modes are in agreement with the charmonium 
result, but most of them tend to lower $\stwob_{\rm eff}$ values so 
that their average appears to be somewhat low.

With the significant decrease in the statistical errors,
the \B-meson Factories enter the era of precision \CP-violation 
measurements in loop processes. 

%
%

\section*{Acknowledgements}

{\small
I am indebted to J\'er\^ome Charles and Zoltan Ligeti 
for informative discussions. I thank my \babar colleagues
for their support, and congratulate the ICHEP'04 committees for
organizing a tremendously interesting conference.
}


\begin{thebibliography}{99}

\bibitem{sin2bbabar}   	\babar\ Collaboration (B.~Aubert \ea), \jprl{89}, 201802 (2002).

\bibitem{sin2bbelle}	Belle Collaboration (K. Abe \ea), \jprd{66}, 071102 (2002).

\bibitem{NewPhys1} 	Y.~Grossman and M.P.~Worah, \plb{395}, 241 (1997).

\bibitem{NewPhys2}	M. Ciuchini \ea, \jprl{79}, 978 (1997);
			D.~London and A.~Soni, \plb{407}, 61 (1997).

\bibitem{grossman}	See, \eg, Y.~Grossman, 
			Int. J. Mod. Phys. A {\bfseries 19}, 907 (2004).

\bibitem{phases}	A.B.~Carter and A.I.~Sanda, \jprd{23}, 1567 (1981);
			I.I.~Bigi and A.I.~Sanda, \npb{193}, 85 (1981);
			R.~Fleischer, Int. J. Mod. Phys. A {\bfseries 12}, 2459 (1997);
			D.~London and A.~Soni, \plb{407}, 61 (1997).

\bibitem{BN}            M.~Beneke and M.~Neubert,
                        \npb{675}, 333 (2003).

\bibitem{ligetiquinn}   Y.~Grossman, Z.~Ligeti, Y.~Nir and H.~Quinn,
                        \jprd{68}, 015004 (2003);
			M.~Gronau, J.L.~Rosner, J.~Zupan, \plb{596}, 107 (2004).

\bibitem{bib:babarNim}	\babar\ Collaboration (B.~Aubert \ea), \nima{479}, 1 (2002).

\bibitem{babarall}	\babar\ Collaboration (B.~Aubert \ea),
			BABAR-CONF-04/33 [hep-ex/0408072], 
			BABAR-CONF-04/40 [hep-ex/0408090], 
			BABAR-CONF-04/19 [hep-ex/0408095], 
			BABAR-CONF-04/25 [hep-ex/0408076], 
			BABAR-CONF-04/30 [hep-ex/0408062] (2004).

\bibitem{newtag}	\babar\ Collaboration (B.~Aubert \ea),
			BABAR-CONF-04/38 [hep-ex/0408127] (2004).

\bibitem{splots}	M.~Pivk and F.~Le Diberder,
			LAL 04-07, physics/0402083 (2004).

\bibitem{belleiso}	Belle Collaboration (K. Abe \ea),
			\jprd{69}, 012001 (2004).

\bibitem{rfit}		A.~H\"ocker, S.~Laplace, H.~Lacker and F.R.~Le Diberder,
			\epjc{21}, 225 (2001).

\bibitem{bellespeng}	Belle Collaboration (K. Abe \ea),
			BELLE-CONF-0435 [hep-ex/0409049].

\bibitem{fzeroks}	\babar\ Collaboration (B.~Aubert \ea),
			BABAR-PUB-04-017 [hep-ex/0406040] (2004).

\bibitem{pizeroks}	\babar\ Collaboration (B.~Aubert \ea),
			BABAR-PUB-04-005 [hep-ex/0403001] (2004).

\end{thebibliography}
\end{document}